\documentclass[showpacs,twocolumn,prl,amsmath,amssymb]{revtex4}

\usepackage{graphicx}
\usepackage{dcolumn}
\usepackage{bm}
\begin{document}
\title{
Electron-phonon interaction in cuprate-oxide superconductors}
\author{Fusayoshi J. Ohkawa}
\affiliation{Division of Physics, Graduate School of
Science,  Hokkaido University, Sapporo 060-0810, Japan}
\email{fohkawa@phys.sci.hokudai.ac.jp}
\received{?? December 2003}
\begin{abstract} 
We propose a novel electron-phonon interaction arising from the modulation of the
superexchange interaction by phonons. It is enhanced by spin
and superconducting fluctuations, which are developed mainly because of the
superexchange interaction. It must be responsible for the softening of phonons and
kinks in the dispersion relation of quasi-particles. However, the superexchange
interaction must be mainly responsible for the formation of Cooper pairs.
\end{abstract}
\maketitle

It is an important issue to elucidate the mechanism of high-T$_c$
superconductivity occurring on CuO$_2$ planes \cite{bednorz}.  Two observations,
the softening of phonons \cite{ McQ1,Pint1,McQ2,Pint2,Braden} and kinks in the
quasi-particle dispersion \cite{ Campuzano,lanzara,Johnson,sato}, imply the
relevance of the electron-phonon interaction. One may argue that it must be
responsible for  high-T$_c$ superconductivity \cite{lanzara}. Its origin and
role should be clarified. 

Doped {\em holes} mainly go into O ions. This implies
that the charge susceptibility of $3d$ electrons on Cu ions is much smaller
than that of $2p$ electrons on O ions and charge fluctuation of $3d$ electrons
can never be developed. Then, the conventional electron-phonon interaction, which
directly couples with charge fluctuations, can play no crucial role. 
On the other hand, antiferromagnetic spin (AFS) and superconducting (SC)
fluctuations are certainly developed. One may argue that their developments are
because of the superexchange interaction. It is shown in early papers
\cite{OhSC1,OhSC2} that the condensation of $d\gamma$-wave Cooper pairs  bound
by the superexchange interaction can explain observed $T_c$.  It is shown in a
previous paper \cite{OhPseudo} that pseudo-gaps appear because of large
life-time widths of quasi-particles due to SC fluctuations.  Suggested by these
arguments, we propose in this Letter an electron-phonon interaction arising from
the modulation of the superexchange interaction by phonons.

It is shown in another previous paper \cite{oh-slave} that 
Gutzwiller's quasi-particle band \cite{gutzwiller} lies between the lower and upper
Hubbard bands \cite{hubbard} in metallic phases in the vicinity of the Mott-Hubbard
transition. The superexchange interaction arises from the virtual exchange of pair
excitations of electrons in spin channels across the lower and upper Hubbard bands;
as long as the Hubbard splitting is significant, it works between Gutzwiller's
quasi-particles  or their renormalized ones \cite{OhSupJ}.  When  nonzero
bandwidths of the lower and upper Hubbard bands are ignored,  the exchange constant
between nearest-neighbor Cu ions is given by 
\begin{equation}
J =-  \frac{4V^4}{(\epsilon_d+U-\epsilon_p)^2}
\left[\frac{1}{\epsilon_d+U-\epsilon_p}+\frac{1}{U} \right] ,
\end{equation}
with $V$ the transfer integral between  $3d$ and $2p$ orbits
on adjacent Cu and O ions,
and $\epsilon_d$ and $\epsilon_p$ their energy levels.
The exchange constant 
 depends on $V$,
$\epsilon_p$ and $\epsilon_d$'s of adjacent Cu ions in such a way that
\begin{eqnarray}
\Delta J &=&
\frac{V^4}{(\epsilon_d+U-\epsilon_p)^3}
\left[\frac{6}{\epsilon_d+U-v_p}+\frac{4}{U} \right]
\nonumber\\  && \hspace*{1cm} \times 
(\Delta \epsilon_{di}+\Delta \epsilon_{dj}-2\Delta \epsilon_{p[ij]})
\nonumber \\ 
&& +
2\frac{J}{V} (\Delta V_{i,[ij]}+ \Delta V_{j,[ij]}) ,
\end{eqnarray}
with $\Delta \epsilon_{di}$ a variation of $\epsilon_d$ of the $i$th Cu
ion, $\Delta \epsilon_{p[ij]}$ a variation of $\epsilon_p$ of the $[ij]$th O ion
that lies between the $i$th and $j$th Cu ions, and  $\Delta V_{i,[ij]}$ that of $V$
between the $i$th Cu ion and the $[ij]$th O ion. When we take the $x$- and $y$-axes
along Cu-O-Cu bonds, they  are given by
\begin{subequations}\label{EqVar}
\begin{eqnarray} 
&& \hspace*{-1cm}
\Delta \epsilon_{di} = A_d \left[
{\bf e}_{x}\!\cdot\!({\bf u}_{i,x_+} \!\!-\! {\bf u}_{i,x_-})
\!+\! {\bf e}_{y}\!\cdot\!({\bf u}_{i,y_+} \!\!-\! {\bf u}_{i,y_-})
\right], 
\\ && \hspace*{1cm}
\Delta \epsilon_{p[ij]} = A_p \left[
{\bf e}_{ij}\cdot({\bf u}_{i}-{\bf u}_{j}) \right],
\\ &&
\Delta V_{i,[ij]}+ \Delta V_{j,[ij]}
= A_V \left[
{\bf e}_{ij}\cdot({\bf u}_{i}-{\bf u}_{j}) \right] ,
\end{eqnarray}
\end{subequations}
with $A_d$, $A_p$ and $A_V$ being constants, 
${\bf u}_i$  the displacement of the $i$th Cu ion, 
${\bf u}_{i,\xi_s}$ that of an O ion on the adjacent $s=+$ or $s=-$
side along the $\xi$-axis of the $i$th Cu ion,  ${\bf e}_x =(1,0)$,  ${\bf e}_y
=(0,1)$, and ${\bf e}_{ij}=({\bf R}_i -{\bf R}_j)/|{\bf R}_i -{\bf R}_j|$,
with ${\bf R}_i$ the position of the $i$th Cu ion.
Displacements of the $i$th Cu and the $[ij]$th O ions are given by
\begin{subequations}\label{EqDisp}
\begin{eqnarray}\label{EqDispCu}
&& \hspace*{-0.5cm} 
{\bf u}_i = \sum_{\lambda{\bf q}}
 \frac{\hbar v_{d,\lambda{\bf q}} } 
{\sqrt{ 2NM_d \omega_{\lambda{\bf q}}} } 
e^{i{\bf q}\cdot{\bf R}_i}
{\bm \epsilon}_{\lambda{\bf q}} \left(
b_{\lambda-{\bf q}}^\dag \!+\! b_{\lambda{\bf q}} \right),
\\   \label{EqDispO}
&& \hspace*{-0.5cm} 
{\bf u}_{[ij]} = \sum_{\lambda{\bf q}}
\frac{\hbar v_{p,\lambda{\bf q}}}
{\sqrt{2N M_p \omega_{\lambda{\bf q}}} } 
 e^{i{\bf q}\cdot {\bf R}_{[ij]} }
{\bm \epsilon}_{\lambda{\bf q}} \! \left(
b_{\lambda-{\bf q}}^\dag \!\!+\! b_{\lambda{\bf q}} \right), \quad 
\end{eqnarray}
\end{subequations}
with ${\bf R}_{[ij]} \!=\! ({\bf R}_i \!+\! {\bf R}_j)/2$, 
$M_d$ the mass of Cu ions, $M_p$ the mass of O ions, $b_{\lambda{\bf q}}$ and 
$b_{\lambda-{\bf q}}^\dag$  annihilation and creation operators of phonons with 
polarization $\lambda$ and wave vector ${\bf q}$, $\omega_{\lambda{\bf q}}$ 
energies of phonons,  ${\bm \epsilon}_{\lambda{\bf q}}$ unit polarization
vectors, and $N$ the number of unit cells. The ${\bf q}$ dependence of
$v_{d,\lambda{\bf q}}$ and $v_{p,\lambda{\bf q}}$ can play a crucial role; 
$v_{d,\lambda{\bf q}}=0$ for the breathing modes that bring no changes in
adjacent Cu-Cu distances while $v_{p,\lambda{\bf q}} \simeq 1$ for such modes.

The electronic part can be  well described by
the $t$-$J$ model on a square lattice:
${\cal H} = - \sum_{ij\sigma} t_{ij} d_{i\sigma}^\dag d_{j\sigma} 
\!-\!  (J/2)
\sum_{\left<ij\right>} ({\bf s}_i \cdot {\bf s}_j)   
\!+\! U_{\infty} \! \sum_{i} n_{i\uparrow}n_{i\uparrow} $,
with the summation over $\left<ij\right>$ restricted to nearest
neighbors,
${\bf s}_i = (1/2)\sum_{\alpha\beta}  \left(
\sigma_x^{\alpha\beta} \!, \sigma_y^{\alpha\beta} \!,
\sigma_z^{\alpha\beta} \right) d_{i\alpha}^\dagger d_{i\beta}$,
with $\sigma_x$, $\sigma_y$ and $\sigma_z$ the Pauli matrixes, 
and $n_{i\sigma}= d_{i\sigma}^\dag d_{i\sigma}$. 
An infinitely large on-site repulsion, 
$U_\infty/|t_{\left<ij\right>}|\rightarrow +\infty$, 
is introduced to exclude any doubly occupied sites. 
According to Eq.~(\ref{EqVar}), there are two types of electron-phonon
interactions. When only longitudinal phonons are considered or when
${\bm \epsilon}_{{\lambda}{\bf q}} = (q_x,q_y,q_z) /q $ is assumed, they are
given by
\begin{subequations}\label{EqElPh}
\begin{eqnarray}\label{EqElPhP}
{\cal H}_p &=&
i C_p \sum_{{\bf q}^\prime\bf q}
\frac{\hbar v_{p,\lambda{\bf q}}\bar{\eta}_{s}({\bf q})}
{\sqrt{2 N M_p \omega_{\lambda{\bf q}}}} 
\sum_{\Gamma=s,d} \!
\eta_{\Gamma}(\mbox{$\frac{1}{2}{\bf q}$}) 
 \eta_{\Gamma}({\bf q}^\prime) 
\nonumber \\ &&  \times
\left(b_{\lambda-{\bf q}}^\dag \!+\! b_{\lambda{\bf q}} \right)
\!\left[
{\bf s}\left({\bf q}^\prime\!\!+\!\mbox{$\frac{1}{2}$}{\bf q}\right) \cdot
{\bf s}\left(-{\bf q}^\prime \!\!+\! \mbox{$\frac{1}{2}$}{\bf q}\right)
\right] , 
\\    \label{EqElPhD}
{\cal H}_d &=&
i C_d \sum_{{\bf q}^\prime\bf q}
\frac{\hbar v_{d,\lambda{\bf q}}}{\sqrt{2 N M_d \omega_{\lambda{\bf q}}}} 
\sum_{\Gamma=s,d} 
\bar{\eta}_{\Gamma}({\bf q}) \eta_{\Gamma}({\bf q}^\prime) 
\nonumber \\ &&  \times
\left(b_{\lambda-{\bf q}}^\dag \!+\! b_{\lambda{\bf q}} \right)
\!\left[
{\bf s}\left({\bf q}^\prime \!\!+\! \mbox{$\frac{1}{2}$}{\bf q}\right) \cdot
{\bf s}\left(-{\bf q}^\prime \!\!+\! \mbox{$\frac{1}{2}$}{\bf q}\right)
\right] , \qquad 
\end{eqnarray}
\end{subequations} 
with 
${\bf s}({\bf q}) = (1/\sqrt{N})\!\sum_{\bf k\alpha\beta} 
(1/2) {\bm \sigma}^{\alpha\beta} d_{({\bf k} +\frac{1}{2}{\bf q}) \alpha}^\dag  
d_{({\bf k} -\frac{1}{2}{\bf q}) \beta} $,
\begin{subequations}
\begin{eqnarray}
&& \hspace*{-0.8cm}
C_p =  \frac{4  A_d V^4}{(\epsilon_d+U-\epsilon_p)^3}
\left[\frac{3}{\epsilon_d+U-\epsilon_p}+\frac{2}{U} \right], \quad
\\ && \hspace*{-0.8cm}
C_d = - \frac{2A_pV^4}{(\epsilon_d \!+\! U \!-\! \epsilon_p)^3}
\left[\frac{3}{\epsilon_d \!+\! U \!-\! \epsilon_p} \!+\! \frac{2}{U} \right]
\!+\! \frac{ A_V J}{V}, 
\end{eqnarray}
\end{subequations} 
 %
$\bar{\eta}_{s}({\bf q})$  $=$ 
$2\left[(q_x/q) \sin\left(q_x a/2\right) 
\!+\! (q_y/q) \sin\left(q_y a/2\right)\right] $,
$\bar{\eta}_{d}({\bf q})$ $=$ 
$2\left[(q_x/q) \sin\left(q_x a/2\right) 
- (q_y/q) \sin\left(q_y a/2\right)\right] $, 
$\eta_{s}({\bf k})$  $=$ $\cos(k_xa) + \cos(k_ya)$, and
$\eta_{d}({\bf k})$  $=$ $\cos(k_xa) - \cos(k_ya)$,
with $a$ the lattice constant.

We follow the previous paper \cite{OhPseudo} to treat the infinitely large
$U_\infty$, 
 where a theory of Kondo lattice is developed.
A renormalized single-site approximation (SSA), which includes not only all the
single-site terms but also the Fock term $\Delta\Sigma({\bf k})$ due to the
superexchange interaction, is reduced to solving the Anderson model.
The self-energy of the Anderson model is expanded as
$\tilde{\Sigma}_\sigma(i\varepsilon_n)=
\tilde{\Sigma}(0) + (1-\tilde{\phi}_\gamma) i\varepsilon_n
+\sum_{\sigma^\prime}(1-\tilde{\phi}_{\sigma\sigma^\prime}) 
\Delta\mu_{\sigma^\prime} + \cdots$,
with $\Delta\mu_{\sigma}$
a small spin-dependent chemical potential shift. Note that 
$\tilde{\phi}_\gamma=\tilde{\phi}_{\sigma\sigma}$.
The Wilson ratio is defined by
$\tilde{W}_s = \tilde{\phi}_s/\tilde{\phi}_\gamma$,
with $\tilde{\phi}_s=
\tilde{\phi}_{\sigma\sigma}-\tilde{\phi}_{\sigma-\sigma}$.
For almost half filling, charge fluctuations are suppressed so that
$\tilde{\phi}_c=
\tilde{\phi}_{\sigma\sigma}+\tilde{\phi}_{\sigma-\sigma} \ll 1$.
For such filling, $\tilde{\phi}_\gamma \gg 1$ so that
$\tilde{\phi}_s \simeq 2\tilde{\phi}_\gamma$ or $\tilde{W}_s \simeq 2$. The
dispersion relation of quasi-particles is given by 
$\xi({\bf k}) = 
(1/\tilde{\phi}_\gamma)
\bigl[-\sum_{j}t_{ij} e^{i{\bf k}\cdot
\left({\bf R}_i-{\bf R}_j\right) }
+ \tilde{\Sigma}(0) + \Delta\Sigma({\bf k}) - \mu \bigr]$,
with  $\mu$ the chemical potential. 

The irreducible polarization function $\pi_s(i\omega_l,{\bf q})$ in spin
channels is divided into single-site  
$\tilde{\pi}_s(i\omega_l)$  and multi-site 
$\Delta\pi_s(i\omega_l, {\bf q})$.
The spin susceptibility is given by
$\chi_s(i\omega_l,{\bf q}) =
2 \pi_s(i\omega_l,{\bf q})/\left\{
1 - \left[ \frac{1}{2}J({\bf q}) + 
U_\infty \right]\pi_s(i\omega_l,{\bf q})\right\}$, 
with $J({\bf q})=2J \eta_s({\bf q})$. In Kondo
lattices, local spin fluctuations at different sites interact with each
other by an exchange interaction.  Following this physical picture, we define
an exchange interaction
$I_s(i\omega_l, {\bf q})$ by 
\begin{equation}\label{EqKondoSus}
\chi_s(i\omega_l, {\bf q}) =
\tilde{\chi}_s(i\omega_l)/ 
\left[1 - \mbox{$\frac{1}{4}$}I_s(i\omega_l, {\bf q})
\tilde{\chi}_s(i\omega_l)\right] ,
\end{equation}
with $\tilde{\chi}_s(i\omega_l) \!=\! 2 \tilde{\pi}_s(i\omega_l)/
\left[1 \!-\! U_\infty \tilde{\pi}_s(i\omega_l) \right]$
the susceptibility for the Anderson model. 
Then, 
%
$I_s (i\omega_l, {\bf q}) = J({\bf q}) + 
2 U_\infty^2 \Delta\pi_s(i\omega_l, {\bf q}) $.
When the Ward relation \cite{ward} is made use of,
the irreducible single-site three-point vertex function in spin channels, 
$\tilde{\lambda}_s(i\varepsilon_n,i\varepsilon_n \!+\! i\omega_l;i\omega_l)$,
is given by
\begin{equation}\label{EqThreeL}
\lim_{U_\infty \rightarrow +\infty} U_\infty 
\tilde{\lambda}_s(i\varepsilon_n,i\varepsilon_n+i\omega_l;i\omega_l)
= 2\tilde{\phi}_s / \tilde{\chi}_s(i\omega_l)  ,
\end{equation}
for $|\varepsilon_n| \rightarrow +0$ and 
$|\omega_l| \rightarrow +0$.
 We approximately use Eq.~(\ref{EqThreeL}) for 
$|\varepsilon_n| \alt k_BT_K$ and
$|\omega_l| \alt k_BT_K$, with $T_K$ the Kondo temperature defined by 
$k_BT_K = \left[1/\tilde{\chi}_s(0)\right]_{T\rightarrow 0}$.  
The main term of  $2 U_\infty^2 
\Delta\pi_s(i\omega_l, {\bf q}) $  is an exchange interaction arising
from the virtual exchange of pair excitations of quasi-particles, which is
less effective than $J({\bf q})$. 
The so called spin-fluctuation mediated interaction, whose single-site term
should be subtracted because it is considered in SSA, is given by 
$\frac{1}{4} \!\bigl[
2\tilde{\phi}_s / \tilde{\chi}_s(i\omega_l)\bigr]^2 \!
\bigl[ \chi_s(i\omega_l, {\bf q}) \!-\! 
\tilde{\chi}_s(i\omega_l) \bigr]$.
It is simply given by  
$\tilde{\phi}_s^2 \frac{1}{4} I_s^* (i\omega_l, {\bf q}) $,
with
\begin{equation}\label{EqIs*2}
\frac{1}{4} I_s^*(i\omega_l, {\bf q}) =
\frac{ \frac{1}{4}
I_s (i\omega_l, {\bf q}) }
{1 - \frac{1}{4}I_s(i\omega_l, {\bf q})
\tilde{\chi}_s(i\omega_l) } .
\end{equation}
 Because of these equations, we call $I_s(i\omega_l, {\bf q})$ a 
{\em bare} exchange interaction, $I_s^*(i\omega_l, {\bf q})$ an
 enhanced one, and $\tilde{\phi}_s$ an effective three-point
vertex function in spin channels.

The enhanced one is  expanded as
$I_s^*(i\omega_l,{\bf q}) = I_0^* + 2 I_1^* 
\eta_s({\bf q}) + \cdots$.
The nearest-neighbor $I_1^*$ is mainly responsible for the development of
not only SC but also charge bond-order (CBO) fluctuations \cite{ComCBO}. 
Because contributions from $|\omega_l|\alt k_BT_K$ are the most effective,
we ignore its energy dependence.
An effective SC  susceptibility, which is
multiplied by $\tilde{\phi}_s^2$, is calculated in the
ladder approximation with respect to $I_1^*$:
\begin{equation}\label{EqSCsus}
\chi_{\Gamma=d}^{\mbox{\tiny (SC)}}(i\omega_l,{\bf q}) =
\frac{2  \tilde{W}_s^2 \pi_{d}^{\mbox{\tiny (SC)}}(i\omega_l,{\bf q})}
{1 + \frac{3}{4} I_1^* \tilde{W}_s^2
\pi_{d}^{\mbox{\tiny (SC)}}(i\omega_l,{\bf q})},
\end{equation}
for $\Gamma=d$ wave, with 
\begin{eqnarray}
\pi^{\mbox{\tiny (SC)}}_{\Gamma}(i\omega_l,{\bf q}) &=&
\frac{k_B T}{N}\sum_{n{\bf k}} 
\eta_{\Gamma}^2({\bf k})
\frac{1}{i\varepsilon_n \!-\! \xi({\bf k} \!+\! \frac{1}{2}{\bf q})}
\nonumber  \\ && \times
\frac{1}{- i\varepsilon_n - i\omega_l-\xi(-{\bf k} + \frac{1}{2}{\bf q})} ,
\end{eqnarray} 
Only $d$-wave SC fluctuation are considered in this Letter.
An effective CBO  susceptibility is similarly given by
\begin{equation}\label{EqCBOsus}
\chi_{\Gamma}^{\mbox{\tiny (CBO)}}(i\omega_l,{\bf q}) =
\frac{2 \tilde{W}_s^2\pi_{\Gamma}^{\mbox{\tiny (CBO)}}(i\omega_l,{\bf q})}
{1 + \frac{3}{4} I_1^* \tilde{W}_s^2
\pi_{\Gamma}^{\mbox{\tiny (CBO)}}(i\omega_l,{\bf q})},
\end{equation}
for $\Gamma=s$, $p$ and $d$ waves, with 
\begin{eqnarray}
\pi^{\mbox{\tiny (CBO)}}_{\Gamma}(i\omega_l,{\bf q}) &=&
- \frac{k_B T}{N}\sum_{n{\bf k}} 
\eta_{\Gamma}^2({\bf k})
\frac{1}{i\varepsilon_n \!-\! \xi({\bf k} \!-\! \frac{1}{2}{\bf q})}
\nonumber \\ && \times
\frac{1}{i\varepsilon_n+i\omega_l-\xi({\bf k} + \frac{1}{2}{\bf q})} .
\end{eqnarray}
The form factors of $p$ waves are defined by
$\eta_{x}({\bf k})  = \sqrt{2}\sin(k_xa)$ and 
$\eta_{y}({\bf k})  = \sqrt{2}\sin(k_ya)$.

\begin{widetext}
A renormalized Green function for
phonons  is given by
$D_\lambda(i\omega_l,{\bf q})=
2 \omega_{\lambda{\bf q}} \big/ \bigl[
(i\omega_l)^2 \!-\! \omega_{{\bf q}\lambda}^2
\!+\! 2 \omega_{\lambda{\bf q}}\Delta\omega_\lambda(i\omega_l,{\bf q})
\bigr]$,
with
$\Delta\omega_\lambda(i\omega_l,{\bf q}) = - 
\left(\hbar^2/2 M_p \omega_{\lambda{\bf q}}\right) 
S (i\omega_l,{\bf q}) $.
Because phonons are renormalized by AFS, SC and CBO fluctuations as well as
pair excitations of quasi-particles in charge channels or 
charge density fluctuations, 
we consider four processes shown in Fig.~\ref{fig}. 
When only the part of
$\Gamma=s$ in Eq.~(\ref{EqElPh}) is considered, it follows that
$S = S_{s}  
+ S_{\mbox{\tiny SC}} +
S_{\mbox{\tiny CBO}}  
+ S_{c} $, with 
\begin{subequations}
\begin{eqnarray}\label{EqS-SDW}
&& 
S_{s}(i\omega_l,{\bf q}) = \frac{3}{4^2}  
Y_{\lambda}^2({\bf q}) 
\frac{k_B T}{N} \sum_{l^\prime {\bf q}^\prime} 
\eta_{s}^2 ({\bf q}^\prime)
\chi_s\!\left(i\omega_l+i\omega_{l^\prime}, 
{\bf q}^\prime + \mbox{$\frac{1}{2}$}{\bf q}\right)
\chi_s\!\left(i\omega_l-i\omega_{l^\prime}, 
-{\bf q}^\prime + \mbox{$\frac{1}{2}$}{\bf q}\right),
\\ &&
S_{\mbox{\tiny SC}}(i\omega_l,{\bf q}) =
\frac{3^2}{4^3} Y_{\lambda}^2({\bf q}) 
\frac{k_B T}{N}  \sum_{l^\prime{\bf q}^\prime} 
\chi_{d}^{\mbox{\tiny SC}}\!\left(i\omega_l+i\omega_{l^\prime}, 
{\bf q}^\prime + \mbox{$\frac{1}{2}$}{\bf q}\right)
\chi_{d}^{\mbox{\tiny SC}}\!\left(i\omega_l-i\omega_{l^\prime}, 
-{\bf q}^\prime + \mbox{$\frac{1}{2}$}{\bf q}\right),
\\ &&
S_{\mbox{\tiny CBO}}(i\omega_l,{\bf q}) =
\frac{3^2}{4^3} Y_{\lambda}^2({\bf q}) 
\sum_{\Gamma}
\frac{k_B T}{N} \sum_{l^\prime{\bf q}^\prime}
\chi_{\Gamma}^{\mbox{\tiny CBO}}\!\left(i\omega_l+i\omega_{l^\prime}, 
{\bf q}^\prime + \mbox{$\frac{1}{2}$}{\bf q}\right)
\chi_{\Gamma}^{\mbox{\tiny CBO}}\!\left(i\omega_l-i\omega_{l^\prime}, 
-{\bf q}^\prime + \mbox{$\frac{1}{2}$}{\bf q}\right),
\\ &&  \label{EqZ}
S_{c}(i\omega_l,{\bf q}) =
- \frac{3^2}{4^2} \tilde{W}_s^4 \frac{k_B T}{N}  \sum_{n\sigma{\bf k}} 
Z^2(i\varepsilon_n,i \omega_l; {\bf k},{\bf q})
\frac{1}{i\varepsilon_n \!-\! \xi({\bf k}) }
\frac{1}{i\varepsilon_n \!+\! i \omega_l \!-\!
\xi({\bf k}\!+\! {\bf q} ) } ,
\end{eqnarray}
\end{subequations}
with  
$Y_{\lambda}({\bf q}) = 
\bar{\eta}_s({\bf q}) \left[C_p v_{p,\lambda{\bf q}} 
\eta_{s} (\mbox{$\frac{1}{2}{\bf q}$})   
+  C_d v_{d,\lambda{\bf q}}\sqrt{M_p/M_d} \right]$.
Here, $Z(i\varepsilon_n,i \omega_l; {\bf k},{\bf q})$ is
a vertex function in the
charge-density channel. It is also enhanced by AFS, SC and CBO fluctuations;  
$Z = Z_{s} + Z_{\mbox{\tiny SC}} + Z_{\mbox{\tiny CBO}}  +\cdots$, with 
\begin{subequations}
\begin{eqnarray} \label{EqZS}
&&
Z_{s}(i\varepsilon_n,i \omega_l; {\bf k},{\bf q}) =
Y_{\lambda}({\bf q})
\frac{k_BT}{N} \sum_{l^\prime{\bf q}^\prime} 
\eta_{s}({\bf q}^\prime) 
\frac{
K_s\left(i\omega_{l^\prime} + i \mbox{$\frac{1}{2}$}\omega_l, 
{\bf q}^\prime + \mbox{$\frac{1}{2}$}{\bf q} \right) 
K_s\left(-i\omega_{l^\prime} + i \mbox{$\frac{1}{2}$}\omega_l, 
-{\bf q}^\prime + \mbox{$\frac{1}{2}$}{\bf q} \right)
}{ 
i\varepsilon_n + i \omega_l^\prime + \mbox{$\frac{1}{2}$}\omega_l
- \xi({\bf k} +  {\bf q}^\prime + \mbox{$\frac{1}{2}$}{\bf q} ) 
} ,
\\ && \label{EqZSC}
Z_{\mbox{\tiny SC}}(i\varepsilon_n,i \omega_l; {\bf k},{\bf q}) =
\frac{1}{2} Y_{\lambda}({\bf q})
\frac{k_BT}{N} \sum_{l^\prime{\bf q}^\prime} 
\eta_{d}\left({\bf k}- \mbox{$\frac1{2}$}{\bf q}^\prime 
+ \mbox{$\frac1{4}$}{\bf q}\right)
\eta_{d}\left({\bf k}- \mbox{$\frac1{2}$}{\bf q}^\prime 
+ \mbox{$\frac{3}{4}$}{\bf q}\right)
\nonumber \\ && \hspace*{4cm} \times 
\frac{
K_{d}^{\mbox{\tiny SC}}
\left(i\omega_{l^\prime} - i \mbox{$\frac{1}{2}$}\omega_l, 
{\bf q}^\prime - \mbox{$\frac{1}{2}$}{\bf q} \right)
K_{d}^{\mbox{\tiny SC}}
\left(i\omega_{l^\prime} + i \mbox{$\frac{1}{2}$}\omega_l, 
{\bf q}^\prime + \mbox{$\frac{1}{2}$}{\bf q} \right)
}{ 
- i\varepsilon_n + i \omega_l^\prime - i \mbox{$\frac{1}{2}$}\omega_l
- \xi(-{\bf k} + {\bf q}^\prime - \mbox{$\frac{1}{2}$}{\bf q} ) },
\\ && \label{EqZCBO}
Z_{\mbox{\tiny CBO}}(i\varepsilon_n,i \omega_l; {\bf k},{\bf q})  =
- \frac{1}{2} Y_{\lambda}({\bf q})
\frac{k_BT}{N} \!\sum_{l^\prime{\bf q}^\prime} 
\sum_{\Gamma} 
\eta_{\Gamma}\left({\bf k}+ \mbox{$\frac1{2}$}{\bf q}^\prime 
+ \mbox{$\frac1{2}$}{\bf q}\right)
\eta_{\Gamma}\left({\bf k}+ \mbox{$\frac1{2}$}{\bf q}^\prime 
+ \mbox{$\frac{3}{4}$}{\bf q}\right)
\nonumber \\ && \hspace*{4cm} \times \frac{
K_{\Gamma}^{\mbox{\tiny CBO}}
\left(i\omega_{l^\prime} + i \mbox{$\frac{1}{2}$}\omega_l, 
{\bf q}^\prime \!+\! \mbox{$\frac{1}{2}$}{\bf q} \right) 
K_{\Gamma}^{\mbox{\tiny CBO}}
\left(-i\omega_{l^\prime} + i \mbox{$\frac{1}{2}$}\omega_l, 
-{\bf q}^\prime + \mbox{$\frac{1}{2}$}{\bf q} \right) 
}{
i\varepsilon_n + i \omega_l^\prime +i  \mbox{$\frac{1}{2}$}\omega_l
- \xi({\bf k} +  {\bf q}^\prime + \mbox{$\frac{1}{2}$}{\bf q} ) },
\end{eqnarray}
\end{subequations}
with
\begin{equation}\label{EqK}
 K_s(i\omega_l,{\bf q}) \!=\! 
\frac{1}{1 \!-\! \frac{1}{4} I(i\omega_l,{\bf q})
\tilde{\chi}_s(i\omega_l)}, 
\hskip5pt
K_{d}^{\mbox{\tiny SC}} (i\omega_l,{\bf q}) \!=\!
\frac{-\frac{3}{4} I_1^* 
\pi_{d}^{\mbox{\tiny SC}}(i\omega_l,{\bf q})}
{1 \!+\! \frac{3}{4} I_1^* 
\pi_{d}^{\mbox{\tiny SC}}(i\omega_l,{\bf q}) },
\hskip5pt  
K_{\Gamma}^{\mbox{\tiny CBO}} (i\omega_l,{\bf q}) \!=\!
\frac{-\frac{3}{4} I_1^* 
\pi_{\Gamma}^{\mbox{\tiny CBO}}(i\omega_l,{\bf q})}
{1 \!+\! \frac{3}{4} I_1^* 
\pi_{\Gamma}^{\mbox{\tiny CBO}}(i\omega_l,{\bf q}) }. 
\end{equation}

\begin{figure*}
\centerline{\hspace*{0.cm}
\includegraphics[width=18.0cm]{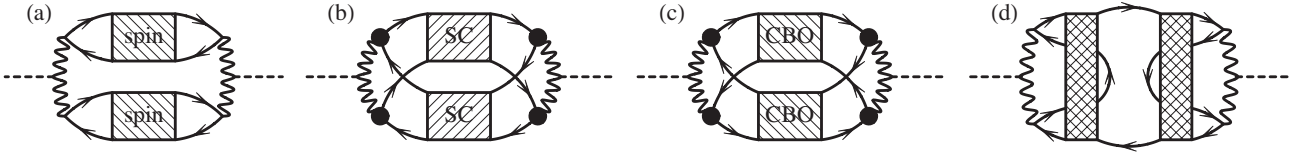}}
\caption[2]{
Four processes contributing to the renormalization of phonons.  
A solid line stands for an electron, a broken line for a phonon, a wavy
line for the superexchange interaction $J$, and a solid circle for
the effective vertex function $\tilde{\phi}_s$. 
 }
\label{fig}
\end{figure*}

No softening occurs for ${\bf q}=0$
because $\bar{\eta}_s({\bf q}\rightarrow 0) \propto |{\bf q}|$. 
When ${\bf q}$ goes from $\Gamma$ point to the zone boundary, the softening must
increase first.  However, it is unlikely that the  softening is the largest at
the zone boundary.
For example, consider the breathing mode at $X$ point,   
${\bf q}_X=(\pm\pi/a,0)$. Because $v_{d,\lambda{\bf q}_X} =0$, 
the electron-phonon interaction described by 
Eq.~(\ref{EqElPhD}) vanishes. 
For ${\bf q}\ne {\bf q}_X$, $v_{d,\lambda{\bf q}}$ is nonzero.
This implies that the softening may not be the largest at $X$ point 
along $\Gamma$-$X$ line. In actual, several experimental data show
that the softening is the largest for ${\bf q}$ a little different from 
${\bf q}_X$ \cite{Pint1,Braden}.

%
Because a low-energy scale is $k_B T_K $, we put
\end{widetext}
\begin{equation}
S(i \omega_l, {\bf q}) \simeq s /k_B T_K ,
\end{equation}
with $s=O(1)$ a dimensionless constant.
It follows that
\begin{equation}
\Delta\omega_\lambda(\omega_{\lambda{\bf q}_X},{\bf q}_X)
\simeq - 15 s c_p^2 \mbox{~meV},  
\end{equation} 
 for
\begin{equation}
C_p \simeq c_p \mbox{~eV}/\mbox{\AA} , \ 
\omega_{\lambda{\bf q}_X}=0.1\mbox{~eV}, \ 
k_BT_K=0.1\mbox{~eV} ,
\end{equation}
with $c_p=O(1)$ a dimensionless constant.
When we take  
\begin{equation}
V\simeq 1.6\mbox{~eV}, \ 
\epsilon_{3d}-\epsilon_{2p}\simeq -1\mbox{~eV}, \  
U\simeq 5\mbox{~eV}, 
\end{equation}
following the previous paper \cite{OhSupJ}, 
it follows that
$c_p=(0.5\mbox{-}1)$
for $A_d=(1\mbox{-}2)$~eV/\AA.
If $sc_p^2 \simeq 1$, the observed softening as large as 
\cite{McQ1,Pint1,McQ2,Pint2,Braden}
\begin{equation}\label{EqObsSoft}
\Delta\omega_\lambda(\omega_{\lambda{\bf q}_X},{\bf q}_X) 
\simeq - 10 \mbox{~meV} , 
\end{equation}
can be explained. 
It should be examined whether $sc_p^2$ is actually as large as 1.

A process corresponding to Fig.~1(d) renormalizes quasi-particles.
The self-energy correction is given by 
\begin{eqnarray}
\frac{1}{\tilde{\phi}_\gamma }\Delta \Sigma (i\varepsilon_n, {\bf k}) 
\!\! &=& \!\!
- \frac{k_B T}{N} \sum_{\lambda l{\bf q}} 
g_\lambda ^2(i\varepsilon_n,i \omega_l;{\bf k},{\bf q})
D_\lambda(i\omega_l,{\bf q})
\nonumber \\ && \times
\frac{1}{i\varepsilon_n+i\omega_l
-\xi({\bf k}+{\bf q})} ,
\end{eqnarray}
with
\begin{equation}\label{EqG}
g_\lambda(i\varepsilon_n,i \omega_l; {\bf k},{\bf q}) =
C_p \frac{\hbar}{\sqrt{2M_p \omega_{\lambda{\bf q}}} }
\frac{3}{4}\tilde{W}_s^2
Z(i\varepsilon_n,i \omega_l; {\bf k},{\bf q}) . 
\end{equation}
It is likely  that the contribution of Fig.~\ref{fig}(d) dominate those
of the other three, Figs.~\ref{fig}(a)--(c).  In such a case, 
\begin{equation}
g_\lambda(i\varepsilon_n,i \omega_l; {\bf k},{\bf q}) \simeq
\sqrt{k_B T_K |\Delta\omega_\lambda(\omega_{\lambda{\bf q}_X},{\bf q}_X)|} .
\end{equation}
When the experimental value (\ref{EqObsSoft}) is used, we obtain 
\begin{equation}
g_\lambda(i\varepsilon_n,i \omega_l; {\bf k},{\bf q}) \simeq
30 \mbox{~meV}.
\end{equation}
This is large enough for optical phonons to cause kinks in the dispersion relation
of quasi-particles. Two types of kinks are observed \cite{sato}. The renormalization
by phonons can explain one type of kinks observed in both normal and SC phases.
However, it cannot explain the other type of kinks observed only in SC phases.

The phonon-mediated pair interaction is given by
$g_\lambda ^2(0, 0;{\bf k},{\bf q})D_\lambda(0,{\bf q})$ or  
$- 2g_\lambda ^2(0, 0;{\bf k},{\bf q})/\omega_{\lambda{\bf q}}$.
The softening of phonons is the largest for ${\bf q}\simeq{\bf q}_X$ along
$\Gamma$-$X$ line. This implies that  the pair interaction by phonons
is attractive between nearest neighbors. The nearest-neighbor part of 
$- 2g_\lambda ^2(0, 0;{\bf k},{\bf q})/\omega_{\lambda{\bf q}} $ should be
included in $\frac{3}{4}I_1^*$. 
According to the argument in this Letter, it follows that 
\begin{equation}
- 2g_\lambda ^2(0, 0;{\bf k},{\bf q}_X)/\omega_{\lambda{\bf q}_X}
\simeq - 20 \mbox{~meV}.
\end{equation}
The phonon-mediated interaction cannot be ignored in cuprate-oxide superconductors. 
However, it is smaller than $\frac{3}{4}|I_1^*| \simeq 100\mbox{~meV}$.
The main Cooper-pair interaction must be the superexchange interaction.

There are two other types of electron-phonon interactions.
Note that $\tilde{\phi}_c$ and 
$1/\tilde{\phi}_\gamma$ are small parameters in the vicinity of the Mott-Hubbard
transition.  The conventional one arising from the modulation of $3d$-electron
levels,  which can directly couples with charge fluctuations, gives renormalization
effects higher order in 
$\tilde{\phi}_c$ and $1/\tilde{\phi}_\gamma$. The interaction
arising from the modulation of $t_{ij}$  gives renormalization effects higher
order in  $1/\tilde{\phi}_\gamma$. Then, we ignore both of them.
On the other hand, what are considered in this Letter are of the order
of $(\tilde{\phi}_c)^0(1/\tilde{\phi}_\gamma)^0$.

In conclusion, we propose the electron-phonon interaction arising from the
modulation of the superexchange interaction by  phonons, which
is only relevant for strongly correlated electron liquids in the vicinity of the
Mott-Hubbard transition.  Its novel property is that it can be enhanced by spin,
superconducting, and charge bond-order fluctuations as well as charge density
fluctuations.  A phenomenological argument where parameters are determined from the
observed softening of phonons implies that the enhanced electron-phonon interaction
is also responsible for kinks in the dispersion relation of quasi-particles in
cuprate-oxide high-$T_c$  superconductors. However, it can never be the main
Cooper-pair interaction.  The main one must be the superexchange interaction. 

This work is supported by a Grant-in-Aid for Scientific Research (C)
No.~13640342 from the Japan Society for the promotion of Science.



\begin{references}
\bibitem{ComCBO}
The order parameter of CBO is
$\sum_\sigma \bigl<a_{i\sigma}^\dag a_{j\sigma}\bigr> $
with $i\ne j$; bond order should not be called anisotropic density wave.
\end{references}
\end{document}